\documentstyle[preprint,tighten,aps]{revtex}
\begin{document} 
\draft
\preprint{IASSNS-HEP-95/107,PUPT-1582}
\date{December 1995}
\title{General Static Spherically Symmetric Black Holes of 
Heterotic String on a Six Torus}
\author{Mirjam Cveti\v c$^1$
\thanks{On sabbatic leave from the University of Pennsylvania.
E-mail address: cvetic@sns.ias.edu}
and Donam Youm$^2$
\thanks{On leave from the University of Pennsylvania. 
E-mail addresses: youm@pupgg.princeton.edu; youm@sns.ias.edu}}
\address{$^1$ School of Natural Science, Institute for Advanced
Study\\ Olden Lane, Princeton, NJ 08540 \\ and \\
$^2$ Physics Department, Joseph Henry Laboratories\\ 
Princeton University \\
Princeton, NJ 08544}
\maketitle
\begin{abstract}
{We present the most general static, spherically symmetric solutions 
of heterotic string compactified on a six-torus that conforms to the
conjectured ``no-hair theorem'', by performing a subset of $O(8,24)$ 
transformations, {\it i.e.}, symmetry 
transformations of the effective three-dimensional action for 
stationary solutions, on the Schwarzschild solution.  
The explicit form of the generating solution is determined by six 
$SO(1,1)\subset O(8,24)$ boosts, with the zero Taub-NUT charge 
constraint imposing one constraint among two boost parameters.
The non-nontrivial scalar fields are the axion-dilaton field and 
the moduli of the two-torus.  The general solution, 
parameterized by {\it unconstrained} 28 magnetic  and 28 
electric charges and the ADM mass compatible with the Bogomol'nyi 
bound, is obtained by imposing on the generating solution
$[SO(6)\times SO(22)]/[SO(4)\times SO(20)]\subset O(6,22)$
($T$-duality) transformation and $SO(2)\subset SL(2,R)$ ($S$-duality) 
transformation, which do not affect the four-dimensional space-time. 
Depending on the range of boost parameters, the non-extreme solutions 
have the space-time of either Schwarzschild or  Reissner-Nordstr\" om 
black hole, while extreme ones have either null (or naked) 
singularity, or the space-time of extreme Reissner-Nordstr\" om 
black hole.}
\end{abstract}

\section{Introduction}

The black holes of the effective four-dimensional, heterotic string 
theory compactified on a six-torus 
\cite{HL,KALL,OTKAL,BANK,KHU,DUFFR,SENBH,HETBH,BEKALL}
\footnote{For an overview see, {\it e.g.},  Ref. \cite{HETBHP} 
and references therein.} 
have been the subject of active research, since they may shed light
on quantum aspects of black holes as well as on the role such 
non-perturbative configurations may play in the full string dynamics.

In the past a class of  such black holes with special charge 
configurations \cite{HL,KALL,OTKAL,BANK,KHU,DUFFR,BEKALL} were 
considered and, in particular, all the electrically charged 
stationary black holes were constructed \cite{SENBH}.  Recently, 
progress has been made in constructing a more general class of 
BPS-saturated \cite{HETBH} and non-extreme \cite{HETBHP} 
{\it dyonic} static, spherically symmetric black holes, 
which contain the previously constructed static spherically 
symmetric configurations as special cases. 
 
Such dyonic black holes  correspond to configurations with 
28 electric and 28 magnetic charges of the $U(1)^{28}$ gauge 
symmetry, which are subject to {\it one constraint}. 
The `generating solution' for  this class of solutions is 
parameterized
\footnote{In Ref. \cite{CTP} it was shown that the BPS-saturated 
generating solution is an exact string solution by proving the 
conformal invariance of the corresponding $\sigma$-model.} 
by {\it two} magnetic and {\it two} electric charges of 
$U(1)^{(1)}_{1,\,M} \times U(1)^{(1)}_{2,\,E} \times 
U(1)^{(2)}_{1,\,M} \times U(1)^{(2)}_{2,\,E}$ gauge group.  
Here the superscripts $(1,2)$ denote the Kaluza-Klein and 
two-form gauge group factors, respectively,  and the subscripts 
$(1,\,M)$  and $(2,\,E)$ denote the magnetic and electric 
charges associated with the gauge factors of the first and second 
compactified coordinates, respectively.  Namely, the electric 
and magnetic charges arise from gauge factors associated with 
{\it different} compactified coordinates.  The explicit form of 
this generating solution for the BPS-saturated \cite{HETBH} and 
non-extreme \cite{HETBHP} configurations was obtained by solving 
directly the corresponding Killing spinor equations and the second 
order Euler-Lagrange equations, respectively. 

The general configurations in this class are then obtained by 
imposing fifty $ [SO(6)\times SO(22)]/[SO(4)\times SO(20)] \subset 
O(6,22)$ symmetry transformations and one $SO(2)\subset SL(2,R)$ 
transformation on the generating solution.  Here, $O(6,22)$ and 
$SL(2,R)$ correspond to the $T$-duality and $S$-duality symmetries 
of the four-dimensional effective action, which do not affect the 
four-dimensional space-time metric.  Thus, along with the four 
charge parameters of the generating solution, the subsequent 
$T$- and $S$-duality transformations introduce 51 new charge 
degrees of freedom.  General solutions in this class are then  
specified by the ADM mass (compatible with the Bogomol'nyi bound) 
and  by  28 electric and 28 magnetic charges which are subject to 
one charge constraint, {\it i.e.}, there is one charge degree of 
freedom missing.   On the other hand the most general, static, 
spherically symmetric configuration, consistent with the conjectured 
``no-hair theorem''
\footnote{The content of the no-hair theorem would ensure that
each black hole type solution would be uniquely specified by 
its mass, angular momentum and a set of conserved (electric and 
magnetic) charges.   Such a theorem has not been proven in general and 
was studied in Ref. \cite{NOHAIR} only within special theories like pure 
Einstein gravity or Einstein-Maxwell gravity and that only for 
configurations with regular horizons, {\it i.e.}, it does not apply to 
configurations with naked singularities or singular horizons.  Nevertheless,  
we assume that the no-hair theorem exists and can be applied also to  
the black hole type solutions of the low energy effective action of  
toroidally compactified heterotic string.  See, for example, Ref. 
\cite{GIBB}, where anti-gravitating black holes with scalar hairs  
were studied within $N=8$ supergravity, which contains scalar fields 
as well as gravity and vector fields.  Since such scalar hairs are, in 
general, functions of $U(1)$ charges and the ADM mass, our position is 
that in the broad sense the no-hair theorem is not violated in this and 
other related cases.  Note also that some of the extreme solutions 
presented here have naked singularities or singular horizons, and thus 
general considerations of the no-hair theorem do not apply to such cases.}
\cite{NOHAIR}, should be parameterized by the ADM mass and 56 
{\it unconstrained} charges.  In order to obtain such configurations 
one has to start with a generating solution, whose charge 
configuration is parameterized by  {\it five} parameters.

The purpose of this paper is to complete the program by obtaining 
the explicit form of the generating solution for all the 
static, spherically symmetric configurations of the effective 
heterotic string theory compactified on a six-torus, compatible 
with the conjectured no-hair theorem.  We shall employ a solution 
generating technique by which new solutions are obtained by  performing 
symmetry transformations on a known solution.  The symmetry 
transformations employed are those of the effective three-dimensional 
action for the stationary solutions of the toroidally compactified 
string.  Related solution generating techniques were used to obtain 
the explicit form of a general class of stationary solutions in 
dilaton-Maxwell-Einstein gravity \cite{SENRBH,GK}, all the static 
spherically symmetric black hole solutions of Abelian Kaluza-Klein 
theory
\footnote{The most general four-dimensional static spherically 
symmetric solutions of the $(4+n)$-dimensional Abelian 
Kaluza-Klein theories, as anticipated in Ref. \cite{BREIT}, 
was obtained \cite{CYALL} by applying a specific subset of 
$SO(2,n)$ symmetry transformations of the three-dimensional 
effective action on Schwarzschild solution.  
The explicit form of the generating solution, parameterized by 
four dyonic charges of $U(1)_1\times U(1)_2$ gauge factors of 
the first and second compactified directions, was 
obtained by applying four $SO(1,1)\subset SO(2,n)$ boosts to the
Schwarzschild solution with zero Taub-NUT constraint imposing 
one constraint on two boost parameters, or equivalently, the four 
charges of the generating solution subjected to one constraint.  
(It was observed in Ref. \cite{GIBB} that such a boost transformation 
induces charges of the Kaluza-Klein $U(1)$ gauge field within the 
context of the five-dimensional Kaluza-Klein theory.) 
The most general solution, specified by the ADM mass and 
{\it unconstrained} $n$-magnetic and $n$-electric charges of 
$U(1)^n$ gauge symmetry, is then obtained by imposing  
$2n-3$ $SO(n)/SO(n-2)\subset SO(n)$ transformations, 
{\it i.e}, target space symmetry transformations of the 
four-dimensional action, on the generating solution.} 
as well as all the electrically charged rotating black hole 
solutions of the heterotic string compactified on a six-torus 
\cite{SENBH}.  The solution generating technique to obtain 
the most general class of static, spherically symmetric solutions 
of heterotic string compactified on a six-torus was already 
anticipated in Ref. \cite{SENBH}.  In this paper we provide an 
explicit realization of the symmetry transformations and an 
explicit form of the generating solution. 

The effective three-dimensional action for  heterotic string  
compactified on a seven-torus was given in Ref. \cite{SENTHREE},  
where it was shown to have a symmetry larger than the direct 
product of the three-dimensional target space $T$-duality 
$O(7,23,R)$ and the four-dimensional $S$-duality $SL(2,R)$.  
Namely, since $SL(2,R)$ symmetry in four-dimensions does not 
commute with $O(7,23)$ symmetry in three-dimensions, these two 
symmetries generate a larger symmetry group $O(8,24)$ in three 
dimensions. 

In order to obtain the explicit form of the general generating 
solution, we shall impose six $SO(1,1)\subset O(8,24)$ boost 
transformations on the Schwarzschild (neutral) black hole solution.  
The three boosts $SO(1,1)\subset O(7,23)$ on the Schwarzschild 
solution generate \cite{SENBH} only electric charges.  
In fact, the  three additional boosts $SO(1,1)\subset 
O(8,24)-O(7,23)$ induce the necessary magnetic charges. The two
`magnetic' and two `electric' boosts  yield the previous  
generating solution with the charge configuration of 
$U(1)^{(1)}_{1,\,M} \times U(1)^{(1)}_{2,\,E} \times 
U(1)^{(2)}_{1,\,M} \times U(1)^{(2)}_{2,\,E}$ gauge group.  
The zero Taub-NUT constraint imposes one constraint among the 
two left-over boosts. Thus, the generating solution is specified 
by a five-parameter charge configuration (associated with the 
first two compactified dimensions), and the non-extremality 
parameter (related to the Schwarzschild black hole mass), which 
parameterizes a deviation of the ADM mass from the supersymmetric 
limit.  The most general solution is then obtained by imposing 
fifty  $[SO(6)\times SO(22)]/[SO(4)\times SO(20)]\subset O(6,22)$ 
(four-dimensional $T$-duality) and one $SO(2)\subset SL(2,R)$ 
($S$-duality) transformations, which do not affect the 
four-dimensional space-time metric. 

In chapter II, we summarize the four-dimensional (chapter IIa) and 
three-dimensional  (chapter IIb) effective actions of  heterotic 
strings on a six-torus and a seven-torus, respectively, and give the 
relationship (chapter IIc) between fields of the
three-dimensional and four-dimensional effective actions 
for the case of the static, spherically symmetric configuration.  
The  solution generating technique with the explicit set of 
symmetry transformations is discussed in chapter IIIa, and the
explicit form of the generating solution for all the static, 
spherically symmetric black holes (compatible with conjectured no-hair 
theorem) of the heterotic string compactified on a six-torus 
is given in chapter IIIb.  The global space-time structures 
and thermal properties of such solutions are discussed in 
chapters IVa and IVb, respectively.  Conclusions are given in 
chapter V.

\section{Action of Heterotic String  on a Six-Torus and a Seven-Torus}

For the purpose of fixing the notations, we first summarize 
(see Refs. \cite{SENTHREE,SENFOUR} and references therein) structure 
of the effective three-dimensional and four-dimensional field theories 
of heterotic string compactified on a seven-torus and a six-torus, 
respectively.  In addition we shall give the relationship between 
the fields of the effective three-dimensional and four-dimensional 
actions for the four-dimensional static, spherically symmetric 
configurations. 

The starting point is the effective field theory of heterotic string
in ten-dimensions, which is described by the $N=1$ supergravity theory 
coupled to $N=1$ super Yang-Mills theory in ten-dimensions.  
At generic points in the moduli space of heterotic string, 
the massless bosonic fields are given by $\hat{G}_{MN}$, 
$\hat{B}_{MN}$, $\hat{A}^I_M$ and $\Phi$ ($0 \le M,N \le 9$, 
$1 \le I \le 16$), which correspond to the graviton, two-form field, 
$U(1)^{16}$ part of the ten-dimensional gauge group, and the 
dilaton field, respectively.  The ten-dimensional action 
of these massless bosonic modes is given by
\begin{equation}
{\cal L} = \sqrt{-\hat{G}}\,[{\cal R}_{\hat{G}} + \hat{G}^{MN}
\partial_M \Phi \partial_N \Phi - {\textstyle {1\over {12}}}
\hat{H}_{MNP}\hat{H}^{MNP}-{\textstyle{1\over 4}}\hat{F}^I_{MN}
\hat{F}^{I\,MN}],
\label{10daction}
\end{equation}
where $\hat G \equiv {\rm det}\,\hat{G}_{MN}$, ${\cal R}_{\hat G}$ is 
the Ricci scalar of the metric $\hat{G}_{MN}$, $\hat{F}^I_{MN} = 
\partial_M \hat{A}^I_N - \partial_N \hat{A}^I_M$ and $\hat{H}_{MNP} = 
\partial_M \hat{B}_{NP}-{1\over 2}\hat{A}^I_M \hat{F}^I_{NP} + 
{\rm cyc.\ perms.}$ are the field strengths of $\hat{A}^I_M$ and 
$\hat{B}_{MN}$, respectively.  We choose the mostly positive signature 
convention $(-++\cdots +)$ for the metric $\hat{G}_{MN}$.

\subsection{Effective Four-Dimensional Action}

We parameterize the solutions in terms of fields in effective 
four-dimensional heterotic string on a six-torus \cite{SENFOUR,SENBH}, 
whose action is obtained by dimensionally reducing (\ref{10daction}) 
on a six-torus down to four-dimensions.  One uses the following 
Kaluza-Klein Ansatz for the metric:
\begin{equation}
\hat{G}_{MN}=\left ( \matrix{e^{2\varphi}g_{{\mu}{\nu}}+
G_{{m}{n}}A^{(1)\,m}_{{\mu}}A^{(1)\,n}_{{\nu}} & A^{(1)\,m}_{{\mu}}
G_{{m}{n}}  \cr  A^{(1)\,n}_{{\nu}}G_{{m}{n}} & G_{{m}{n}}} \right ),
\label{4dkk}
\end{equation}
where $A^{(1)\,m}_{\mu}$ ($\mu = t,r,\theta,\phi$; 
$m=1,...,6$) are four-dimensional Kaluza-Klein $U(1)$ gauge fields 
and $\varphi \equiv \Phi - {1\over 2}{\rm ln}\,{\rm det}\, G_{mn}$ 
is the four-dimensional dilaton field.  The effective 
four-dimensional action is specified by the following massless 
(four-dimensional) bosonic fields: the (Einstein-frame) graviton 
$g_{\mu\nu}$, the complex scalar field (dilaton-axion field)
\footnote{$\Psi$ is the axion which is equivalent to the two-form 
field $B_{\mu\nu}$ through the duality transformation  
$H^{\mu\nu\rho} = -{e^{2\varphi} \over {\sqrt{-g}}}
\varepsilon^{\mu\nu\rho\sigma}\partial_{\sigma}\Psi$.}
$S=\Psi + ie^{-\varphi}$, 28 $U(1)$ gauge fields ${\cal A}^i_{\mu} 
\equiv (A^{(1)\, m}_{\mu}, A^{(2)}_{\mu\, m}, A^{(3)\, I}_{\mu})$ 
defined as $A^{(2)}_{\mu\,m} \equiv \hat{B}_{\mu\,m} + 
\hat{B}_{mn}A^{(1)\,n}_{\mu} + {1\over 2}\hat{A}^I_m 
A^{(3)\,I}_{\mu}$, $A^{(3)\,I}_{\mu} \equiv \hat{A}^I_{\mu} - 
\hat{A}^I_m A^{(1)\,m}_{\mu}$, and the following 
symmetric $O(6,22)$ matrix  of the scalar fields (moduli):
\begin{equation}
M=\left ( \matrix{G^{-1} & -G^{-1}C & -G^{-1}a^T \cr 
-C^T G^{-1} & G + C^T G^{-1}C +a^T a & C^T G^{-1} a^T 
+ a^T \cr -aG^{-1} & aG^{-1}C + a & I + aG^{-1}a^T} 
\right ), 
\label{modulthree}
\end{equation} 
where $G \equiv [\hat{G}_{mn}]$, $C \equiv [{1\over 2}
\hat{A}^{(I)}_{{m}}\hat{A}^{(I)}_{n}+\hat{B}_{mn}]$ and 
$a \equiv [\hat{A}^I_{{m}}]$ are specified in terms of the 
internal parts of ten-dimensional fields.
The effective four-dimensional Lagrangian takes the form 
\footnote{We set $\alpha^{\prime}=2$, the four-dimensional Newton's 
constant $G_N = {\textstyle {1\over 8}} \alpha^{\prime}= 
{1\over 4}$ and the compactification radii $R_{m}=
\sqrt{\alpha^{\prime}}=\sqrt 2$.} 
\cite{MS,SENFOUR}: 
\begin{equation}
{\cal L} = \sqrt{-g}
\left[{\cal  R}_g -{\textstyle {1\over 2}}\partial_{\mu} \varphi 
\partial^{\mu} 
\varphi -{\textstyle{1\over 2}}{\rm e}^{2\varphi}
\partial_{\mu} \Psi \partial^{\mu} \Psi-{\textstyle 
{1\over 4}}{\rm e}^{-\varphi}
{\cal F}^i_{\mu\nu}(LML)_{ij}{\cal F}^{j\,\mu\nu} 
+{\textstyle  {1\over 8}}{\rm Tr}(\partial_{\mu} ML 
\partial^{\mu}ML)\right] ,
\label{string}
\end{equation}
where $g\equiv {\rm det}\,g_{\mu\nu}$, ${\cal R}_g$ is the Ricci 
scalar of $g_{\mu\nu}$, and ${\cal F}^i_{\mu\nu} = \partial_{\mu} 
{\cal A}^i_{\nu}- \partial_{\nu} {\cal A}^i_{\mu}$ are 
the $U(1)^{28}$ gauge field 
strengths.  The four-dimensional  effective action (\ref{string}) 
is invariant under the $O(6,22)$ transformations ($T$-duality) 
\cite{MS,SENFOUR}:
\begin{equation}
M \to \Omega M \Omega^T ,\ \ \ {\cal A}^i_{\mu} \to \Omega_{ij}
{\cal A}^j_{\mu}, \ \ \ g_{\mu\nu} \to g_{\mu\nu}, \ \ \ S \to S ,
\label{tdual}
\end{equation}
where  $\Omega$ is an $O(6,22)$ invariant matrix, {\it i.e.}, with the 
following property:
\begin{equation}
\Omega^T L \Omega = L ,\ \ \ L =\left ( \matrix{0 & I_6& 0\cr
I_6 & 0& 0 \cr 0 & 0 &  I_{16}} \right ),
\label{4dL}
\end{equation}
where $I_n$ denotes the $n\times n$ identity matrix.

In addition, the corresponding equations of motion and Bianchi 
identities are invariant under the $SL(2,R)$ transformations 
($S$-duality) \cite{SENFOUR}:
\begin{equation}
S \to { {{aS+b}\over{cS+d}}},\ \ M\to M ,\ \ g_{\mu\nu}\to 
g_{\mu\nu},\ \ {\cal F}^i_{\mu\nu} \to (c\Psi + d)
{\cal F}^i_{\mu\nu} + ce^{-2\varphi} (ML)_{ij}
\tilde{\cal F}^j_{\mu\nu},
\label{sdual}
\end{equation}
where  $\tilde{\cal F}^{i\,\mu\nu} = {1\over 2\sqrt{-g}} 
\varepsilon^{\mu\nu\rho\sigma}{\cal F}^i_{\rho\sigma}$ and 
$a,b,c,d \in R$ satisfy $ad-bc=1$.
At the quantum level, the parameters of both $T$- and $S$-duality 
transformations become integer-valued, corresponding to the exact 
symmetry of the perturbative string theory and the conjectured 
non-perturbative symmetry of string theory, respectively.

\subsection{Effective Three-Dimensional Action}

The dimensional reduction of the ten-dimensional action 
(\ref{10daction}) on a seven-torus down to three dimensions 
\cite{SENTHREE} can be achieved by the following choice of 
the Kaluza-Klein Ansatz for the metric:
\begin{equation}
\hat{G}_{MN}=\left ( \matrix{e^{2{\bar\varphi}}h_{\bar{\mu}\bar{\nu}}+
\bar{G}_{\bar{m}\bar{n}}\bar{A}^{\bar{m}}_{\bar{\mu}}
\bar{A}^{\bar{n}}_{\bar{\nu}} & \bar{A}^{\bar{m}}_{\bar{\mu}}
\bar{G}_{\bar{m}\bar{n}}  \cr  \bar{A}^{\bar{n}}_{\bar{\nu}}
\bar{G}_{\bar{m}\bar{n}} & \bar{G}_{\bar{m}\bar{n}}} \right ),
\label{3dkk}
\end{equation}
where ${\bar\varphi} \equiv \Phi - {1\over 2}{\rm ln}\,{\rm det}
\,\bar{G}_{\bar{m}\bar{n}}$ ($\bar{m}, \bar{n}=t,1,...,6 \equiv 1,...,7$) 
is the three-dimensional dilaton and $\bar{A}^{\bar{m}}_{\bar{\mu}}$ 
($\bar{\mu}=r,\theta,\phi$) are the three-dimensional Kaluza-Klein 
$U(1)$ gauge fields.  The three-dimensional action for massless 
(three-dimensional) bosonic fields contains the following fields: 
the graviton $h_{\bar{\mu}\bar{\nu}}$, the dilaton ${\bar \varphi}$, 
30 $U(1)$ gauge fields $\bar{{\cal A}}^{\bar{i}}_{\bar{\mu}} \equiv 
(\bar{A}^{\bar{m}}_{\bar{\mu}}, \bar{A}^{7+\bar{m}}_{\bar{\mu}}, 
\bar{A}^{14+I}_{\bar{\mu}})$ defined as  $\bar{A}^{7+\bar{m}}_
{\bar{\mu}} \equiv \hat{B}_{\bar{\mu}\bar{m}} + 
\hat{B}_{\bar{m}\bar{n}}\bar{A}^{\bar{n}}_{\bar{\mu}}+
{1\over 2}\hat{A}^I_{\bar{m}}\bar{A}^{14+I}_{\bar{\mu}}$ 
and $\bar{A}^{14+I}_{\bar{\mu}} \equiv 
\hat{A}^I_{\bar{\mu}} - \hat{A}^I_{\bar{m}}
\bar{A}^{\bar{m}}_{\bar{\mu}}$ with the field strengths 
$\bar{{\cal F}}^{\bar{i}}_{\bar{\mu}\bar{\nu}} \equiv \partial_
{\bar{\mu}}\bar{{\cal A}}^{\bar{i}}_{\bar{\nu}} - \partial_{\bar{\nu}}
\bar{{\cal A}}^{\bar{i}}_{\bar{\mu}}$, the two-form field 
$\bar{B}_{\bar{\mu}\bar{\nu}}$
\footnote{Since in three-dimensions $\bar{B}_{\bar{\mu}\bar{\nu}}$ 
has no physical degrees of freedom, its field strength 
$\bar{H}_{\bar{\mu}\bar{\nu}\bar{\rho}}$ can be set to zero.}, 
and the following symmetric $O(7,23)$ matrix of scalar fields:
\begin{equation}
\bar{M}=\left ( \matrix{\bar{G}^{-1} & -\bar{G}^{-1}\bar{C} & 
-\bar{G}^{-1}\bar{a}^T \cr 
-\bar{C}^T \bar{G}^{-1} & \bar{G} + \bar{C}^T \bar{G}^{-1}\bar{C} +
\bar{a}^T \bar{a} & \bar{C}^T \bar{G}^{-1} \bar{a}^T 
+ \bar{a}^T \cr -\bar{a}\bar{G}^{-1} & \bar{a}\bar{G}^{-1}\bar{C} + 
\bar{a} & I + \bar{a}\bar{G}^{-1}\bar{a}^T} 
\right )
\label{modul1}
\end{equation}
defined in terms of internal parts $\bar{G} \equiv 
[\hat{G}_{\bar{m}\bar{n}}]$, $\bar{C} \equiv [{1\over 2}
\hat{A}^{I}_{\bar{m}}\hat{A}^{I}_{\bar{n}}+
\hat{B}_{\bar{m}\bar{n}}]$ and $\bar{a} \equiv 
[\hat{A}^I_{\bar{m}}]$ of the ten-dimensional fields.

The resulting three-dimensional action is invariant under the 
$O(7,23)$ transformations:
\begin{equation}
\bar{M} \to \bar{\Omega} \bar{M} \bar{\Omega}^T, \ \ \ 
\bar{{\cal A}}^{\bar{i}}_{\bar{\mu}} 
\to \bar{\Omega}_{\bar{i}\bar{j}}\bar{{\cal A}}^{\bar{j}}_{\bar{\mu}},
\ \ \ 
h_{\bar{\mu}\bar{\nu}} \to h_{\bar{\mu}\bar{\nu}}, \ \ \ 
\bar{B}_{\bar{\mu}\bar{\nu}} \to \bar{B}_{\bar{\mu}\bar{\nu}}, \ \ \ 
{\bar\varphi} \to {\bar\varphi} , 
\label{o723}
\end{equation}
where $\bar{\Omega} \in O(7,23)$, {\it i.e.}, it has the 
following property:
\begin{equation}
\bar{\Omega} \bar{L} 
\bar{\Omega}^T = \bar{L}, \ \ \ \bar{L}=\left 
( \matrix{0 & I_7 & 0 \cr I_7 & 0 & 0 \cr 0 & 0 & I_{16}} \right ) . 
\label{3dL}
\end{equation}
At the quantum level $O(7,23)$ becomes integer valued, and is the 
$T$-duality  symmetry group of the perturbative heterotic string 
theory on a seven-torus.

Since in three-dimensions vector fields are dual to scalar fields, 
one can make the following duality transformations to trade the 
three-dimensional $U(1)$ fields $\bar{{\cal A}}^{\bar{i}}_{\bar{\mu}}$ 
with a set of scalar fields $\psi \equiv [\psi^i]$ \cite{SENTHREE}:
\begin{equation}
\sqrt{-h}e^{-2{\bar\varphi}}h^{\bar{\mu}\bar{\mu}^{\prime}}
h^{\bar{\nu}\bar{\nu}^{\prime}}(\bar{M}\bar{L})_{\bar{i}\bar{j}}
\bar{{\cal F}}^{\bar{j}}_{\bar{\mu}^{\prime}\bar{\nu}^{\prime}} = 
{\textstyle {1\over 2}}\epsilon^{\bar{\mu}\bar{\nu}\bar{\rho}}
\partial_{\bar{\rho}}\psi^{\bar{i}}.
\label{dual}
\end{equation}
Then the three-dimensional action reduces to the following 
form \cite{SENTHREE}:
\begin{equation}
{\cal L} = {\textstyle {1\over 4}}\sqrt{-h}\,[{\cal R}_h + 
{\textstyle {1\over 8}}h^{\bar{\mu}\bar{\nu}}{\rm Tr}
(\partial_{\bar{\mu}}{\cal M}{\bf L}
\partial_{\bar{\nu}}{\cal M}{\bf L})],
\label{3daction}
\end{equation}
where $h\equiv {\rm det}\,h_{\bar{\mu}\bar{\nu}}$, ${\cal R}_h$ 
is the Ricci scalar of the three-dimensional metric 
$h_{\bar{\mu}\bar{\nu}}$.  ${\cal M}$ is a symmetric $O(8,24)$ 
matrix  of three-dimensional scalar fields defined as 
\begin{equation}
{\cal M} = \left ( \matrix{\bar{M}-e^{2{\bar\varphi}}\psi\psi^T &
e^{2{\bar\varphi}}\psi & 
\bar{M}\bar{L}\psi-{1\over 2}e^{2{\bar\varphi}}\psi(\psi^T 
\bar{L} \psi) 
\cr 
e^{2{\bar\varphi}}\psi^T & -e^{2{\bar\varphi}} & {1\over 
2}e^{2{\bar\varphi}}\psi^T \bar{L}\psi \cr 
\psi^T \bar{L}\bar{M} - {1\over 2}e^{2{\bar\varphi}}\psi^T
(\psi^T \bar{L}\psi) & 
{1\over 2}e^{2{\bar\varphi}}\psi^T \bar{L}\psi & -e^{-2{\bar\varphi}} 
+ \psi^T \bar{L}\bar{M}\bar{L}\psi - 
{1\over 4}e^{2{\bar\varphi}}(\psi^T \bar{L}\psi)^2}
 \right ). 
\label{modultwo}
\end{equation}
The action is manifestly invariant under the $O(8,24)$ 
transformations:
\begin{equation}
{\cal M} \to {\bf \Omega} {\cal M} {\bf \Omega}^T, \ \ \ \ 
h_{\bar{\mu}\bar{\nu}} \to h_{\bar{\mu}\bar{\nu}}, 
\label{o824}
\end{equation}
where ${\bf \Omega} \in O(8,24)$, {\it i.e.},
\begin{equation}
{\bf \Omega}{\bf L}{\bf\Omega}^T={\bf L } , \ \  \
 {\bf L} = \left (\matrix{\bar{L} & 0 & 0 \cr 0 & 0 & 1 
\cr 0 & 1 & 0}\right )\  .
\label{3dbfo}
\end{equation}

Within the effective field theories described above, we shall be 
interested in four-dimensional static, spherically symmetric 
configurations.  The relationship between such fields of the 
three-dimensional action (\ref{3daction}) and the four-dimensional 
action (\ref{string}) are given in the following subsection. 

\subsection{Relationship Between the Fields of the Three- and 
Four-Dimensional Actions}

For the static, spherically symmetric configurations the 
four-dimensional (Einstein frame) space-time metric has the 
following form:
\begin{equation}
g_{\mu\nu}dx^{\mu}dx^{\nu} = -\lambda(r) dt^2 + \lambda(r)^{-1} dr^2 
+R(r)(d\theta^2 + {\rm sin}^2 \theta d\phi^2), 
\label{4dmetric}
\end{equation}
and the four-dimensional scalar fields in $M$ (cf. 
(\ref{modulthree})), {\it i.e.}, moduli,
and the dilaton-axion field $S=\Psi+i{\rm e}^{-\varphi}$ 
depend on the radial coordinate $r$, only. 
 
The spherical symmetry implies that 28 ${\cal A}^i_\mu$ vector 
potentials have non-zero $t$ and $\phi$ components, whose
asymptotic values  ($r\to \infty$) specify the electric 
and magnetic charges, respectively. The Maxwell's equations and 
Bianchi identities determine the $U(1)$ field strengths to be
\begin{equation}
{\cal F}^i_{\theta\phi} = L_{ij}\beta_j\, {\rm sin}\,\theta,
\ \ \ \ \
{\cal F}^i_{tr} = { {e^{\varphi} \lambda(r)\over r^2}}
M_{ij}(\alpha_j +  \Psi \beta_j), 
\label{elecmag}
\end{equation}
where $\vec \alpha$ and $\vec \beta$ correspond to the conserved  
(quantized) 28 electric and 28 magnetic charge vectors, and $L$ 
is defined in (\ref{4dL}).

The physical  magnetic and electric  charges
\begin{equation}
\vec P\equiv (P^{(1)}_m; P^{(2)}_m; P^{(3)}_I), \ \ \ \ \ \
\vec Q\equiv(Q^{(1)}_m; Q^{(2)}_m; Q^{(3)}_I) ,
\label{ccc}
\end{equation}
are related to the conserved  (quantized) charge vectors 
$\vec \alpha$ and $\vec \beta$  in the following way:
\begin{equation}
P_i=L_{ij}{\beta}_j\ , \ \ \ \ Q_i = e^{\varphi_{\infty}}
M_{ij\,\infty}(\alpha_j + \Psi_{\infty}\beta_j),
\label{charges}
\end{equation}
where the subscript $\infty$ refers to the asymptotic ($r\to\infty$)
values of the corresponding fields. 

The corresponding three-dimensional space metric takes the form:
\begin{equation}
h_{\bar{\mu}\bar{\nu}}dx^{\bar{\mu}}dx^{\bar{\nu}}= 
dr^2+{\bar R}(r)(d\theta^2 + {\rm sin}^2 \theta d\phi^2), 
\label{metric}
\end{equation}
where the four-dimensional metric components are related to 
the three-dimensional ones as ${\bar R}=\lambda R $. 
The three-dimensional scalar fields in $\cal M$ (cf. (\ref{modultwo}))
depend on the radial coordinate $r$, only.
 
For static, spherically symmetric configuration described above, 
the four-dimensional fields are related to the three-dimensional 
fields in the following way:
\begin{eqnarray}
G_{mn}&=&\bar{G}_{1+m,\,1+n},\ \ \ \ \ B_{mn}=\bar{B}_{1+m,\,1+n},
\ \ \ \ \ a^I_m=\bar{a}^I_{1+m},\cr
\varphi&=&\bar{\varphi}+{\textstyle {1\over 2}}[{\rm ln}\,{\rm det}\, 
\bar{G}_{\bar{m}\bar{n}} - {\rm ln}\, {\rm det}\,G_{mn}],\ \ \ 
\lambda=e^{-\varphi}[{\rm det}\,\bar{G}_{\bar{m}\bar{n}}/
{\rm det}\,G_{mn}]^{1\over 2},\ \ \ 
R=\bar{R}/\lambda ,\cr
A^{(1)\,m}_t &=& G^{mn}\bar{G}_{1,1+n},\ \ \ 
A^{(1)\,m}_{\phi}=G^{mn}\bar{A}^{\bar{m}}_{\phi}\bar{G}_{\bar{m},1+n}, 
\ \ \  A^{(3)\,I}_t = \bar{a}^I_1 - a^I_m A^{(1)\,m}_t,\cr 
A^{(3)\,I}_{\phi}&=& \bar{A}^{14+I}_{\phi}+a^I_{\bar{m}}
\bar{A}^{\bar{m}}_{\phi}-a^I_m A^{(1)\,m}_{\phi},\ \ \ 
A^{(2)}_{t\,m}=\bar{B}_{1,1+m}+B_{mn}A^{(1)\,n}_t+{\textstyle 
{1\over 2}}a^I_m A^{(3)\,I}_t, \cr 
A^{(2)}_{\phi\,m}&=&\bar{A}^{8+m}_{\phi}-\bar{B}_{1+m,\bar{n}}
\bar{A}^{\bar{n}}_{\phi}-{\textstyle {1\over 2}}
a^I_m\bar{A}^{14+I}_{\phi}+B_{mn}A^{(1)\,n}_{\phi}+
{\textstyle {1\over 2}}a^I_m A^{(3)\,I}_{\phi}, \cr
B_{\phi t}&=& \bar{A}^{(8)}_t - \bar{B}_{1\bar{n}}\bar{A}^{\bar{n}}_t 
-{\textstyle {1\over 2}}\bar{a}^I_1 \bar{A}^{(14+I)}_t +
{\textstyle {1\over 2}}(A^{(1)\,m}_{\phi}A^{(2)}_{t\,m}
-A^{(1)\,m}_{t}A^{(2)}_{\phi\,m})-A^{(1)\,m}_{\phi}
B_{mn}A^{(1)\,n}_t .
\label{3d4drel}
\end{eqnarray}

\section{Solution}

In this chapter we first spell out the solution generating technique, 
and the explicit sequence of symmetry transformations imposed on the 
Schwarzschild solution.  Then we give the explicit form of the 
generating solution.  We further  spell out the $T$- and $S$-duality
transformations on the generating solution, thus yielding the most general
solution specified by 28 electric and 28 magnetic charges and the ADM mass.
Finally, we compare the special cases of this general solution 
with existing examples in the literature.

\subsection{Solution Generating Technique}

An arbitrary asymptotic value of the scalar field matrix ${\cal M}$ 
can be transformed into the form 
\begin{equation}
{\cal M}_{\infty} = {\rm diag}(-1,\overbrace{1,...,1}^6,-1,
\overbrace{1,...,1}^{22},-1,-1) \equiv I_{4,28}
\label{asympt}
\end{equation}
by imposing an $O(8,24)$ transformation, {\it i.e.}, 
${\cal M}_{\infty} \to {\bf \Omega} {\cal M}_{\infty}{\bf 
\Omega}^T = I_{4,28}$ (${\bf \Omega} \in O(8,24)$).  Thus, 
without loss of generality we can confine the analysis by 
choosing the asymptotic value of the form ${\cal M}_\infty=
I_{4,28}$ and, then, obtain the solutions with arbitrary 
value of ${\cal M}_\infty$ by undoing the above constant 
$O(8,24)$ transformation.  Then, the subset of $O(8,24)$ 
transformations that preserve this asymptotic value for 
$\cal M$ is $SO(8) \times SO(24)$
\footnote{Similarly, the arbitrary asymptotic values of the 
four-dimensional moduli matrix $M$ and the dilaton-axion $S$ 
can be brought into the form $M_{\infty}=I_{28}$ and $S_{\infty}=i$ 
by using constant $O(6,22)$ and $SL(2,R)$ transformations,
respectively.  The subsets of four-dimensional symmetry transformations
which preserve these asymptotic values are $SO(6)\times SO(22)$ and 
$SO(2)$, respectively.}.   

The starting point of generating the static, spherically symmetric 
solutions with the most general charge configurations is the following 
three-dimensional form of the Schwarzschild black hole solution
\footnote{To obtain the most general rotating charged black hole 
solution one imposes the {\it same} subsets of $O(8,24)$ 
transformations on the Kerr solution.  For the sake of simplicity 
we confined ourselves to the static solutions, only.}:  
\begin{equation}
{\cal M} = {\rm diag}(-{\textstyle {r\over{r-m}}},
\overbrace{1,...,1}^{6},-{\textstyle {{r-m}\over r}},
\overbrace{1,...,1}^{22},-{\textstyle {r\over{r-m}}},
-{\textstyle {{r-m}\over r}}), 
\label{schwarz}
\end{equation}
and the three dimensional space metric (\ref{metric}) is specified by 
${\bar R}=r(r-m)$.  Here, $m$ is the ADM mass of the Schwarzschild 
black hole. 

The subset of $O(8,24)$ transformations that generate new solutions 
from (\ref{schwarz}) is the quotient space $[O(22,2)\times O(6,2)]/
[O(22)\times O(6) \times SO(2)]$ \cite{SENBH}, where $SO(2)$ is the 
subset of $SL(2,R)$ transformation in four-dimensions that preserves 
the asymptotic value $S_{\infty}=i$ as discussed in the previous 
footnote.  The quotient space is parameterized by 57 parameters.  
One has to notice that in imposing a subset of $O(8,24)$ 
transformations on the Schwarzschild solution, in general, 
one induces the unphysical Taub-NUT charge.  To remove unphysical 
Taub-NUT charge one has to impose one constraint among the 
parameters of $O(8,24)$ transformations.   Therefore, these 
subsets of $O(8,24)$ transformations, subject to zero Taub-NUT 
charge constraint, introduce $56$ new charge degrees of freedom, 
which correspond to unconstrained $28$ magnetic and $28$ electric 
charges.  Those are the necessary charge degrees of freedom 
specifying the most general static black hole solutions of  
heterotic string on a six-torus, compatible with the conjectured 
no-hair theorem.

We shall proceed with the following sequence of the symmetry 
transformations.  In order to obtain the explicit form of 
the generating solution we shall apply six necessary 
$SO(1,1)\subset O(8,24)$ boost transformations on the 
Schwarzschild solution, which would yield, after imposing the zero 
Taub-NUT constraint, the generating solution specified by five 
parameter charge configurations.   We then impose  on the generating  
solution the $[SO(6)\times SO(22)]/[SO(4)\times SO(20)]$ and $SO(2)$ 
transformations, {\it i.e.}, the subsets of $T$- and $S$-duality 
transformations, which do not affect the four-dimensional space-time.  
Thus, one in turn obtains the most general configuration 
specified by the mass and {\it unconstrained} 28 electric and
28 magnetic charges of $U(1)^{28}$ gauge group. 

\subsection{Explicit Form of the Generating Solution}

For the purpose of obtaining the explicit form of the most general 
solution, it is convenient to first impose four successive $SO(1,1)$ 
boosts ${\bf \Omega}_{p1,p2,q1,q2}$ on the Schwarzschild solution 
(\ref{schwarz}) with boost parameters $\delta_{p1}, \delta_{p2}, 
\delta_{q1}, \delta_{q2}$, which generate non-extreme 
$U(1)^{(1)}_{1,\,M} \times U(1)^{(1)}_{2,\,E} \times 
U(1)^{(2)}_{1,\,M} \times U(1)^{(2)}_{2,E}$ solution \cite{HETBHP}. 
The boost transformation ${\bf \Omega}_{p1}$ has the following  form:
\begin{equation}
{\bf \Omega}_{p1}\equiv\left (\matrix{1&\cdot&\cdot&\cdot&\cdot&\cdot&
\cdot\cr
\cdot& {\rm cosh}\delta_{p1}&\cdot&\cdot&\cdot&\cdot&{\rm sinh}
\delta_{p1}\cr 
\cdot&\cdot&I_6&\cdot&\cdot&\cdot&\cdot \cr
\cdot&\cdot&\cdot&{\rm cosh}\delta_{p1}&\cdot&-{\rm sinh}
\delta_{p1}&\cdot\cr
\cdot&\cdot&\cdot&\cdot&I_{21}&\cdot&\cdot\cr
\cdot&\cdot&\cdot&-{\rm sinh}\delta_{p1}&\cdot&{\rm cosh}
\delta_{p1}&\cdot\cr
\cdot&{\rm sinh}\delta_{p1}&\cdot&\cdot&\cdot&\cdot&{\rm cosh}
\delta_{p1}}\right ), 
\label{boost}
\end{equation}
where the dots denote the corresponding zero entries and 
${\bf \Omega}_{p2}$ has the analogous form with the non-trivial 
entries (with positive
signs in front of sinh's) in the $\{9th, 32nd\}$  and (with negative
signs in front of sinh's) in the  $\{2nd, 31st\}$, columns and rows.    
${\bf\Omega}_{q1}$ [${\bf\Omega}_{q2}$] has the non-trivial entries  
(with positive signs in front of sinh's) in the $\{8th, 10th\}$ 
[$\{1st, 10th\}$] and (with negative signs in front of sinh's) 
in the $\{1st, 3rd\}$, [$\{3rd, 8th\}$] columns and rows.

This solution has among 28 magnetic ${\vec P}$ and 28 electric 
${\vec Q}$ charges (cf. (\ref{ccc})) only four non-zero
charges of the  Kaluza-Klein and two-form  gauge fields, which 
are associated with the first two compactified directions, only.  
The two magnetic and the two electric charges arise from  
{\it different} compactified directions.  Namely:
\begin{eqnarray}
P^{(1)}_1 &=& m{\rm sinh}\delta_{p1}{\rm cosh}\delta_{p1} \equiv P_1,
\ \ \ \ \, Q^{(1)}_1 = 0, \cr
P^{(1)}_2 &=& 0, \ \ \ \ \ \ \ \ \ \ \ \ \ \ \ \ \ \ \ 
\ \ \ \ \ \ \ \ \ \ \ \
Q^{(1)}_2 = m{\rm cosh}\delta_{q1}{\rm sinh}\delta_{q1} \equiv 
Q_1,\cr 
P^{(2)}_1 &=& m{\rm sinh}\delta_{p2}{\rm cosh}\delta_{p2}\equiv 
P_2,\ \  
Q^{(2)}_1 = 0, \cr
P^{(2)}_2 &=& 0, \ \ \ \ \ \ \ \ \ \ \ \ \ \ \ \ \ \ \ 
\ \ \ \ \ \ \ \ \ \ \ \
Q^{(2)}_2 = m{\rm cosh}\delta_{q2}{\rm sinh}\delta_{q2} \equiv Q_2, 
\label{charge}
\end{eqnarray}
and the ADM mass given by
\begin{equation}
M_{ADM} =\hat{P}_1 + \hat{P}_2 + \hat{Q}_1 + \hat{Q}_2, 
\label{mass}
\end{equation}
where the hated quantities are defined as $\hat{P}_1 \equiv \pm 
\sqrt{(P_1)^2 + \beta^2}$, etc.  The signs $\pm$ in the 
definitions of the hated quantities are determined \cite{HETBHP} 
by the requirement that the above ADM 
mass formula approaches the Bogomol'nyi bound:
\begin{equation}
M_{BPS}= |P_1+P_2|+|Q_1+Q_2|,
\label{bpsmass}
\end{equation}
as $\beta \to 0$.  Here $\beta \equiv  m/2 > 0$ is the non-extremality 
parameter that measures a deviation from the supersymmetric limit. 
In the extreme (supersymmetric) limit ($\beta\to 0$), one takes
$\delta_{p1,p2,q1,q2}\to \infty$ in a way that $\beta{\rm
e}^{2\delta_{p1,p2,q1,q2}}$  become finite parameters, in order for
charges $P_{1,2},Q_{1,2}$  to remain non-zero.
Note, that when the relative signs of the $P_{1,2}$ (and $Q_{1,2}$) 
parameters are opposite the corresponding non-extreme solutions are 
{\it ultra-extreme} \cite{HETBHP}, {\it i.e.}, their ADM masses 
{\it are not} compatible with the Bogomol'nyi bound (their ADM 
masses are smaller than  the BPS mass), and thus are not in the 
spectrum of states.  

We shall concentrate on non-extreme (regular) solutions, 
{\it i.e.}, those which reduce to the BPS solutions with the 
same relative signs for $P_{1,2}$ (and $Q_{1,2}$) in the 
supersymmetric limit, and consequently $\hat P_1=+\sqrt{(P_1)^2+
\beta^2}$, etc
\footnote{There are also hybrid solutions \cite{HETBHP} with 
the opposite relative signs for one type of charges, say, 
magnetic ones with $|P_1|>|P_2|$, and the same relative 
signs for the other type of charges, {\it i.e.}, electric ones.  
These solutions are singular, however, the non-extreme solutions  
satisfy $M_{ADM}\ge M_{BPS}$ and, therefore, are in the spectrum, 
provided $\sqrt{(P_1)^2+\beta^2}\left(1/\sqrt{( P_2)^2+\beta^2}+1/ 
\sqrt{( Q_1)^2+\beta^2}+1/\sqrt{( Q_2)^2+\beta^2}\right) \ge 1$.  
This constraint can be satisfied only for a restricted range of 
$\beta$ and charge parameters.}. 

The explicit form of the above solutions was given in Ref. 
\cite{HETBHP} by directly solving the Euler-Lagrange equations. 
Here, the same solution is obtained by the solution 
generating technique specified by the four boost parameters 
$\delta_{p1,p2,q1,q2}$ and $m$, or equivalently by four charge 
parameters $P_1, P_2, Q_1, Q_2$ and the non-extremality parameter 
$\beta$.  This solution now provides an intermediate step on the 
way to obtain the generating solution with the charge configuration 
specified by five parameters and the non-extremality parameter.   

The missing one more charge degree of freedom can be obtained by 
performing on the above solution two more $SO(1,1)\subset O(8,24)$ 
transformations, parameterized by two boost parameters 
$\delta_{1,2}$.  The corresponding ${\bf \Omega}_{1}$ 
[${\bf\Omega}_2$] has the nontrivial entries  (with positive signs 
in sinh's) in the $\{1st, 2nd\}$ [$\{3rd, 31st\}$] and  (with 
negative signs in sinh's) in  the $\{8th, 9th\}$ [$\{10th, 32nd\}$] 
columns and rows.

The above two boosts do introduce two new charges, {\it i.e.},
$P^{(2)}_2$ and $Q^{(1)}_1$, as well as the (unphysical) Taub-NUT 
charge
\footnote{Note that without loss of generality, one 
could have chosen other sets of two boosts, which would 
induce other pairs of $U(1)$ charges, {\it i.e.},
($P^{(1)}_2,Q^{(2)}_1$), ($P^{(1)}_2,P^{(2)}_2$), etc.  
All such solutions are related to the above generating 
solution through  $SO(2)\times SO(2)\subset O(2,2)$ 
and $SO(2)\subset SL(2,R)$ transformations, {\it i.e.}, subsets 
of the two-torus $T$-duality and $S$-duality transformations, which 
do not affect the four-dimensional space-time.}.  
The Taub-NUT charge can be eliminated by imposing the following 
constraint on the boost parameters $\delta_{1,2}$:
\begin{equation}
P_1{\rm tanh}\delta_1 -  Q_2{\rm tan}\delta_2 = 0.  
\label{notaubnut}
\end{equation}

Without loss of generality, we assume that $Q_2 \geq P_1$ and then 
$\delta_2$ is expressed in terms of $\delta_1$ as
\footnote{For the case $Q_2 \leq P_1$, the role of the boost parameters
$\delta_1 $ and $\delta_2$ are interchanged.}:
\begin{equation}
{\rm cosh}\,\delta_2 = Q_2 {\rm cosh}\,\delta_1 /\Delta, \ \ \ \ \ 
{\rm sinh}\,\delta_2 = P_1 {\rm sinh}\,\delta_1 /\Delta,
\label{boostrel}
\end{equation}
where $\Delta \equiv {\rm sign}(Q_2) \sqrt{(Q_2)^2 {\rm cosh}^2 
\delta_1 - (P_1)^2 {\rm sinh}^2 \delta_1}$.    

The final form of the generating solution (with zero Taub-NUT charge),
expressed in terms of the four-dimensional fields, as specified in 
Section IIa and related to the fields of the three-dimensional action 
through (\ref{metric}) and (\ref{3d4drel}), can be written in the 
following form
\footnote{The generating solution which saturates the 
corresponding Bogomol'nyi bound (BPS-saturated solution), 
{\it i.e.}, the case with $\beta=0$ and $\delta_1$ finite, was also
found  in Ref. \cite{CT} as (an exact string) 
solution of the conformal invariance constraints for the 
$\sigma$-model which corresponds to the 
chiral-null model with the non-trivial four-dimensional transverse 
part.  There, the generating solution was parameterized by (five) 
non-zero charge parameters $P^{(1,2)}_1\equiv P_{1,2}$, 
$Q^{(1,2)}_2\equiv Q_{1,2}$ and $Q^{(1)}_1=-Q^{(2)}_1\equiv q$.  
These two  (BPS-saturated) generating solutions are
related to each another by subsets of $SO(2)\times SO(2)\subset
O(2,2)$ (two-torus $T$-duality)  and  $SO(2)\subset SL(2,R)$ 
($S$-duality) transformations.}:  
\begin{eqnarray}
\lambda &=& { {{(r+\beta)(r-\beta)}\over 
{(XY-Z^2)^{1\over 2}}}},\ \ \ 
R=(XY-Z^2)^{1\over 2},\ \ \ \ \ \ \  
e^{2\varphi}= { {W^2 \over {XY-Z^2}}},
\cr 
\partial_r \Psi &=& { {1\over {\Delta^3\,W}}}\left[
\Delta^2 (P_1 Q_1 + P_2 Q_2)+ P_1 Q_2 [(P_1)^2(r+\hat{Q_2})-
(Q_2)^2(r+\hat{P}_1)] \right.\cr
&\times&\left.{{{P_1 P_2 (r-\hat{Q}_1){\rm sinh}^2 \delta_1 
+ Q_1 Q_2 (r+\hat{P}_2){\rm cosh}^2 \delta_1 } \over {XY-Z^2}}}
\right]{{\rm sinh}\delta_1 
{\rm cosh}\delta_1}, 
\cr 
G_{11}&=&{{X\over {(r+\hat{P}_1)(r+\hat{Q}_2)}}}, \ \ \ 
G_{22}={ {Y\over {(r+\hat{P}_1)(r+\hat{Q}_2)}}}, \ \ \
G_{12}=-{ {Z\over {(r+\hat{P}_1)(r+\hat{Q}_2)}}}, 
\cr 
B_{12}&=&-{ {{[(Q_2)^2(r+\hat{P}_1)-(P_1)^2(r+
\hat{Q}_2)]{\rm cosh}\delta_1 {\rm sinh}\delta_1} \over 
{\Delta (r+\hat{P}_1)(r+\hat{Q}_2)}}},\cr
G_{ij}&=&\delta_{ij}, \  \ \  B_{ij}=0, \ \ \ (i,j \neq 1, 2),
\ \ \ a^I_m=0\
\label{gensol}
\end{eqnarray}
with  
\begin{eqnarray}
X&=&r^2+[(\hat{P}_2 +\hat{Q}_2){\rm cosh}^2 \delta_1 
+(\hat{Q}_1 - \hat{P}_1){\rm sinh}^2 \delta_1]r +
(\hat{P}_1\hat{Q}_1{\rm sinh}^2 \delta_1 
+ \hat{Q}_2 \hat{P}_2{\rm cosh}^2 \delta_1),\cr
Y&=&r^2+{\textstyle {1\over \Delta^2}}(P_1)^2 (\hat{P}_2 -\hat{Q}_2)
{\rm sinh}^2 \delta_1 + (Q_2)^2 (\hat{P}_1 +\hat{Q}_1)
{\rm cosh}^2 \delta_1]r \cr
&+& {\textstyle{1\over \Delta^2}}[(P_1)^2 \hat{Q}_2 
\hat{P}_2\sinh^2\delta_1 +(Q_2)^2\hat{Q}_1\hat{P}_1 
{\rm cosh}^2 \delta_1],\cr
Z&=&{\textstyle {1\over \Delta}}[(P_1 P_2 +
Q_1 Q_2)r +(\hat{P}_1 Q_1 Q_2 + 
\hat{Q}_2 P_1 P_2)]{{\rm cosh}\delta_1 {\rm sinh}\delta_1},\cr
W&=& r^2 +{\textstyle {1\over \Delta^2}}[(Q_2)^2 (\hat{P}_1 +
\hat{P}_2){\rm cosh}^2 \delta_1 + (P_1)^2 (\hat{Q}_1 - \hat{Q}_2)
{\rm sinh}^2 \delta_1]r\cr
&+& {\textstyle {1\over \Delta^2}}[(Q_2)^2\hat{P}_1
\hat{P}_2 {\rm cosh}^2 \delta_1 + (P_1)^2 \hat{Q}_1
\hat{Q}_2 {\rm sinh}^2 \delta_1], 
\label{def}
\end{eqnarray}
where again $\Delta \equiv {\rm sign}(Q_2) \sqrt{(Q_2)^2 
{\rm cosh}^2 \delta_1 - (P_1)^2 {\rm sinh}^2 \delta_1}$.    

For the sake of simplicity of the above expressions, the radial 
coordinate $r$ is chosen so that the outer horizon is at $r=\beta$.  
This solution has the following non-zero charges:
\begin{eqnarray}
P^{(1)}_1 &=& P_1 Q_2/\Delta, \ \ \ \ \ \ \ \ \ \ \ \ \ \ \ \ 
\ \ \ \ \ \ \, 
Q^{(1)}_1 = (\hat{P}_1 - \hat{P}_2 - \hat{Q}_1 - \hat{Q}_2)
{\rm cosh}\delta_1 {\rm sinh}\delta_1,\cr 
P^{(1)}_2 &=& 0, \ \  \ \ \ \ \ \ \ \ \ \ \ \ \ \ \ \ \ \ \ \ \ 
\ \ \ \ \ \ \ \ \
Q^{(1)}_2 = (Q_1 Q_2 {\rm cosh}^2 \delta_1 + P_1 P_2 {\rm sinh}^2 
\delta_1)/\Delta, \cr
P^{(2)}_1 &=& (Q_2 P_2 {\rm cosh}^2 \delta_1 + 
 Q_1 P_1 {\rm sinh}^2 \delta_1)/\Delta, 
\ \ \ \ \ \ \ \ \ \ \ \ \ \ \ \ \ \ Q^{(2)}_1 = 0, \cr
P^{(2)}_2&=& P_1 Q_2 
(Q_2-Q_1-P_1-P_2){\rm sinh}\delta_1{\rm cosh}\delta_1 /
\Delta^2, \ \ \ \ \ 
Q^{(2)}_2 = \Delta,
\label{ch}
\end{eqnarray}
and the ADM mass formula, compatible with the  BPS bound
\cite{HETBH,DUFFtrial}, is given by:
\begin{eqnarray}
M_{ADM}&=&{\textstyle {1\over \Delta^2}}
[(P_1)^2(\hat{P}_2-\hat{Q}_2){\rm sinh}^2 \delta_1
+ (Q_2)^2(\hat{P}_1 + \hat{Q}_1){\rm cosh}^2 \delta_1]\cr
&+&(\hat{P}_2+\hat{Q}_2){\rm cosh}^2 \delta_1 + 
(\hat{Q}_1-\hat{P}_1){\rm sinh}^2 \delta_1. 
\label{adm}
\end{eqnarray}
Note that the solution is specified by three electric and three 
magnetic charges (subject to zero Taub-Nut constraint) associated 
with the $U(1)^{(1)}_1 \times U(1)^{(2)}_1 \times U(1)^{(1)}_2 
\times U(1)^{(2)}_2$ Kaluza-Klein and the two-form 
fields (the superscripts 1 and 2, respectively) of the first and the 
second compactified dimension (the subscripts 1 and 2, respectively).  
The solution is therefore parameterized by {\it five} parameters 
$P_1, P_2, Q_1, Q_2$ and $\delta_1$ as well as the non-extremality 
parameter $\beta$.  Alternatively, the configuration is specified by
six charges (subject to one constraint) and the mass, compatible with 
the Bogomol'nyi bound.  At this point, it seems to be difficult to 
express the generating solution in terms of physical 
parameters, such as physical charges and the ADM mass.

Among the scalar fields (cf. (\ref{gensol})), both the dilaton 
$\varphi$ and the axion $\Psi$ of the complex scalar field $S$ 
vary with $r$, and only the moduli of the two-torus ($G_{11},
G_{22},G_{12},B_{12}$) in the moduli $M$ (cf. (\ref{modulthree})) 
vary with $r$.  Note that in the limit $\delta_1\to 0$ the 
solutions (\ref{gensol}) reduce to the generating solution with 
four charge parameters (\ref{charge}), only.  In this case, the 
axion and the off-diagonal toroidal moduli $G_{12}$ and $B_{12}$ 
are turned off.

\subsection{$S$- and $T$-Duality Transformations}

The additional 51 charge degrees of freedom necessary in 
parameterizing the most general static, spherically 
symmetric solutions are obtained through the subset 
of $O(6,22)$ ($T$-duality) and $SL(2,R)$ 
($S$-duality) transformations as stated before.  
The expressions for the charges after the transformations 
are given by:
\begin{equation}
\vec{\cal Q}^{\prime} = {1\over \sqrt{2}} {\cal U}^T \left ( \matrix{U_6 
({\bf e}_u-{\bf e}_d)\cr U_{22}\left(\matrix{{\bf e}_u+{\bf e}_d \cr
0_{16}}\right)}\right ), 
\ \ \ \ \ \ 
\vec{\cal P}^{\prime} = {1\over \sqrt{2}} {\cal U}^T \left (\matrix{U_6 
({\bf m}_u-{\bf m}_d)\cr U_{22}\left(\matrix{{\bf m}_u+{\bf m}_d \cr 0_{16}}
\right)}\right), 
\label{totch}
\end{equation}
where 
\begin{eqnarray}
{\bf e}^T_u &\equiv& (Q^{(1)}_1 {\rm cos}\gamma + 
P^{(2)}_1{\rm sin}\gamma , Q^{(1)}_2 {\rm cos}\gamma + 
P^{(2)}_2{\rm sin}\gamma ,\overbrace{0,...,0}^{4}),\ 
{\bf e}^T_d \equiv (P^{(1)}_1{\rm sin}\gamma , Q^{(2)}_2 
{\rm cos}\gamma , \overbrace{0,...,0}^{4}),\cr 
{\bf m}^T_u &\equiv& (P^{(1)}_1{\rm cos}\gamma , Q^{(2)}_2 
{\rm sin}\gamma , \overbrace{0,...,0}^{4}),\ 
{\bf m}^T_d \equiv (P^{(2)}_1 {\rm cos}\gamma + 
Q^{(1)}_1{\rm sin}\gamma , P^{(2)}_2 {\rm cos}\gamma + 
Q^{(1)}_2{\rm sin}\gamma ,\overbrace{0,...,0}^{4}), 
\label{submat}
\end{eqnarray}
$\gamma$ is the $SO(2) \subset SL(2,R)$ rotational angle, 
$U_6 \in SO(6)$, $U_{22} \in SO(22)$, $0_{16}$ is a 
$(16\times 1)$-matrix with zero entries, and 
${\cal U} \equiv {1\over \sqrt{2}}\left ( \matrix{I_6&-I_6&0\cr 
I_6&I_6&0\cr 0&0&I_{16}} \right )$.  The complex scalar $S$ 
and the moduli field $M$ are transformed to
\begin{equation}
S^{\prime}= {{(\Psi{\rm cos}\gamma - {\rm sin}\gamma)+ie^{-\varphi}
{\rm cos}\gamma}\over {(\Psi{\rm sin}\gamma + {\rm cos}\gamma)+ie^{-\varphi}
{\rm sin}\gamma}}, \ \ \ \ \ \ \ 
M^{\prime}={\cal U}^T \left(\matrix{U_6 &0 \cr 0 & U_{22}}\right){\cal U}
M{\cal U}^T \left (\matrix{U^T_6 & 0 \cr 0 & U^T_{22}}\right){\cal U},
\label{genscalar}
\end{equation}
where $\Psi$, $e^{-\varphi}$ and $M$ are the axion, the dilaton and 
the moduli field of the generating solutions (\ref{gensol}).  
Then, the most general ``string frame'' metric $g^{string}_{\mu\nu}$ 
is obtained through conformal transformation of the four-dimensional 
metric given by $g^{string}_{\mu\nu}= g_{\mu\nu}/{\rm Im}\,(S)$, where 
$g_{\mu\nu}={\rm diag}(-\lambda(r),\lambda(r)^{-1},R(r),{\rm sin}^2 
\theta R(r))$ is the ``Einstein frame'' metric for the generating 
solutions given in (\ref{gensol}).   

Since these transformations do not affect the 
four-dimensional ``Einstein-frame'' space-time metric, global 
space-time and thermal properties of the general solution are 
parameterized in terms of the six parameters of the generating solution.  
The study of the global space-time of such solutions is the topic of 
the next chapter.

In principle, one should be able to write the general solution, 
specified by 28 electric and 28 magnetic charges and the ADM mass, 
in terms of $T$- and $S$-duality invariant quantities.   We shall 
not attempt to do that in this paper
\footnote{For the BPS-saturated  solutions, the ADM mass
\cite{HETBH,DUFFtrial,CT} as well as the area of the horizon 
(for the regular solutions) \cite{CT} was cast in the manifestly 
$T$- and $S$-duality invariant form.}.

\subsection{Special Examples of the General Solution}

The above general solution, {\it i.e.}, the generating solution 
accompanied with $S$- and $T$-duality transformations, contains as 
a proper subset the previously known spherically symmetric black 
holes in the heterotic string theory.  Examples, known in the literature,
correspond to the cases when $\delta_1=0$ (Cf., Ref. \cite{HETBHP}), 
and special choices of the charges $P_{1,2}$ and $Q_{1,2}$.  
We conclude this section by illustrating  these special cases.
\begin{itemize}
\item Spherically symmetric black hole solutions in heterotic string theory
with different dilaton-gauge coupling $a$ \cite{GM,GHS,HW}:\\
$P_1=P_2=Q_1=Q_2\neq 0$ case reduces to the Reissner-Nordstr\"om 
black hole solution, {\it i.e.} $a=0$; 
when three charges are non-zero and equal, the solution is 
that of $a=1/\sqrt{3}$ case \cite{DUFFtrial}; when  only two magnetic charges 
[or only two electric charges] are non-zero and equal, the solution is 
 that of $a=1$ case \cite{KALL}; when only one charge is non-zero, 
the solution reduces to that of $a=\sqrt{3}$ case, which contains in the
extreme limit ($\beta \to 0$) as subsets 
($i$) $P_1 \neq 0$ case: the Kaluza-Klein monopole solution of Gross 
and Perry, and Sorkin \cite{KKMON}, and ($ii$) $P_2 \neq 0$ case: 
H-monopole solution \cite{KHU}.   

\item $P_1=P_2$ and $Q_1=Q_2$ solutions with imposed subsets 
of $S$- and $T$-duality  transformations correspond to a general 
class of axion-dilaton black hole solutions found by Kallosh and 
Ortin \cite{OTKAL}.

\item The solutions with $Q_1$ and $Q_2$ non-zero, when supplemented 
by $S$- and $T$-duality transformations, correspond to a  general class of 
electrically charged black holes in heterotic string  constructed by Sen 
\cite{SENBH}.  The $S$-duality counterpart of such general class of solutions
is the purely magnetically charged solutions found by Behrndt and Kallosh 
\cite{BKAL}.

\item The {\it non-supersymmetric extreme solutions}, {\it i.e.}, 
those with $\beta\to 0$, $P_2=Q_1=0$, $|Q_2|-|P_1|\to 0$ and 
$\delta_1 \to\infty$, while keeping $\beta e^{2|\delta_1|}$ and 
$(|Q_2|-|P_1|)e^{2|\delta_1|}$ as finite parameters, reduce to 
the extreme dyonic five-dimensional Kaluza-Klein black hole studied 
in Ref. \cite{GW}, after $S$- and $T$-duality transformations are imposed. 

\end{itemize}

\section{Global Space-Time Structure and Thermal Properties of 
the Solutions}

We shall now classify all the possible global space-time and thermal 
properties of the obtained solutions.
 Since the subsets of 
$O(6,22)$ and $SL(2,R)$ transformations leave the four-dimensional 
(Einstein frame) space-time metric invariant, it is sufficient to 
consider, for that purpose, the generating solution specified by 
(\ref{gensol}).  Therefore, all the possible space-time properties 
of the solutions are determined by six parameters $P_{1,2}$, 
$Q_{1,2}$, $\delta_1$ and the non-extremality parameter $\beta$.
In the following we first discuss the global space-time properties, 
and then the thermal properties of the solutions.  We separate 
the solutions into non-extreme, {\it i.e.}, when  $\beta >0$, and 
extreme ones, {\it i.e.}, when $\beta =0$.  Within each class we 
then analyze their properties according to the range of five 
parameters
\footnote{In the following analysis it is understood that 
$Q_2 \neq 0$, always when $\delta_1 \neq 0$.  
In the case when $Q_2=0$, $P_1$ has to be zero 
automatically due to our initial assumption that 
$|Q_2| \geq |P_1|$.  Then, the zero Taub-Nut constraint 
(\ref{notaubnut}) does not restrict the values of $\delta_{1,2}$. 
In this case, we have non-extreme four-parameter 
solution with non-zero charges given by 
$P^{(2)}_1, P^{(2)}_2, Q^{(1)}_1$ and $Q^{(1)}_2$, equivalently 
parameterized by $Q_1$, $P_2$, and $\delta_{1,2}$.  
Such solutions can be related to the generating solutions 
through the subsets of $SO(2)\times SO(2) \subset O(2,2) 
\subset O(6,22)$ and $SO(2) \subset SL(2,R)$ transformations.}  
$P_{1,2}$, $Q_{1,2}$ and $\delta_1$. 

\subsection{Global Space-Time Structure}

The singularity structure can be explored by studying the 
explicit form (\ref{gensol}) of the four-dimensional metric 
(\ref{4dmetric}).  The Ricci scalar curvature blows up at 
the point ($r=r_{sing}$) where $R=0$, thus corresponding to 
the space-time singularity.  The horizon(s) form  at the point(s) 
($r=r_{hor\, \pm}$) where $\lambda=0$, provided $r_{hor\, \pm} 
> r_{sing}$.  We shall now explore the global 
space-time properties for the two classes of solutions.\\
\\
($i$) {\it Non-extreme solutions} ($\beta > 0$)\\
\\

Recall that we shall consider only the case where both pairs 
($P_1,P_2$) and ($Q_1,Q_2$) have the same relative signs 
(and consequently $\hat P_1=+\sqrt{P_1^2+\beta^2}$, etc. with 
plus signs in the hated quantities), 
since only for this case the ADM mass is compatible with 
the Bogomol'nyi bound for any values of $\beta$
\footnote{In the case when only one of the pairs has opposite 
sign, the ADM mass is compatible with the Bogomol'nyi bound 
for the restricted range of $\beta$ \cite{HETBHP}, only.  
(See also the footnote 9.)  When both of the pairs have the opposite 
signs, the ADM mass is always less than the ADM mass of the 
corresponding supersymmetric solution for any $\beta >0$.}.      
By analyzing the roots of $XY-Z^2$ (cf. (\ref{gensol}),
(\ref{def})), one can see that the singularity always takes 
place at $r_{sing} \leq -\beta$.  On the other hand the zero(s) of 
$\lambda$, corresponding in principle to the outer and/or inner 
horizons, are at $r_{hor\, +}=\beta$ and/or $r_{hor\, -}=-\beta$, 
respectively.  Thus, the global space time of non-extreme 
solutions is that of non-extreme Reissner-Nordstr\"om black 
holes when $r_{sing}<-\beta$, and that of Schwarzschild black 
hole when $r_{sing}=-\beta$. 

The single root of  $XY-Z^2$ at $r_{sing}=-\beta$ takes 
place, in which case the singularity and the inner horizon 
coincide at $r=-\beta$, when 
\noindent{($a$) $\delta_1 \neq 0$ and  $P_1=0$, or}
\noindent{($b$) $\delta_1 =0$ and least one of $P_{1,2}$ and 
$Q_{1,2}$ is zero}. 
On the other hand,
the double root of $XY-Z^2$ at $r_{sing}=-\beta$ takes place, 
in which case the inner horizon disappears and the singularity 
forms at $r=-\beta$, when 
\noindent{($a$) $\delta_1 \neq 0$ and only $Q_2$ is  non-zero, or} 
\noindent{($b$) $\delta_1 = 0$ and at least two of $P_{1,2},Q_{1,2}$ 
are zero.}\\
\\
($ii$) {\it Extreme solutions} ($\beta \to 0$) \\

When the boost parameter  $\delta_1$ is finite, these
solutions saturate the Bogomol'nyi bound, {\it i.e.}, they 
correspond to {\it supersymmetric extreme solutions}.

When both pairs of $(P_1,P_2)$ and $(Q_1,Q_2)$ have the 
same relative signs, the singularity always takes place at 
$r_{sing}\leq 0$.  The inner and outer horizons 
coincide  at $r_{hor\,+}=r_{hor\, -}=0$. Global space-time of such 
supersymmetric extreme solutions is therefore that of the 
extreme Reissner-Nordstr\"om black holes when $r_{sing}<0$, or
the singularity and the horizon coincide when
$r_{sing}=r_{hor\,\pm}=0$.  The latter case happens 
when at least one out of $P_{1,2}, Q_{1}$ (and $Q_2$) parameters 
is zero with $\delta_1 \neq 0$ (with $\delta_1 =0$).  The horizon at 
$r_{hor}=0$ disappears when only $Q_2$ is non-zero with 
$\delta_1 \neq 0$ (or when only one out of $P_{1,2},Q_{1,2}$ 
parameters is non-zero with $\delta_1 =0$).

In the case of at least one of the pairs $(P_{1},P_2)$ 
and $(Q_1,Q_2)$ having the opposite relative sign, the singularity 
at $r_{sing}>0$ is naked \cite{BEHR,CYS}.  

There also exist {\it non-supersymmetric extreme solutions}, 
{\it i.e.}, those with $\beta\to 0$ but without  saturating the 
Bogomol'nyi bound.  To obtain such solutions, one sets $P_2=Q_1=0$, 
and takes the limits $\beta\to 0$, $|Q_2|-|P_1|\to 0$ and 
$\delta_1\to \infty$, while keeping $\beta e^{2|\delta_1 |}$ 
and $(|Q_2|-|P_1|){\rm e}^{2|\delta_1|}$ as finite parameters.   
In this case the time-like singularity is at $r_{sing}<r_{hor}=0$, 
{\it i.e.}, the global space-time is that of the extreme 
Reissner-Nordstr\" om black holes.

\subsection{Thermal Properties of the Solutions}

Thermal properties of  the solution (\ref{gensol}) are specified 
by the four-dimensional space-time at the event horizon 
($r=r_{hor\, +}$).  All the non-extreme solutions ($\beta >0$) 
(compatible with the Bogomol'nyi bound) are regular solutions 
with the event horizon at $r_{hor\, +}=\beta$. 

The entropy $S$ of the system, determined as $S = \pi R(r=\beta)$, 
is of the following form
\begin{eqnarray}
S ={\textstyle {\pi\over { |\Delta|}}} & &\left | \left 
[(\hat Q_2+\beta)(\hat P_2+\beta){\rm cosh}^2 \delta_1 + 
(\hat P_1+\beta)(\hat Q_1-\beta){\rm sinh}^2 \delta_1\right ] 
\right.\times \cr
 & &\left [(Q_2)^2(\hat Q_1+\beta)(\hat P_1+\beta) 
{\rm cosh}^2 \delta_1  + (P_1)^2 (\hat Q_2+\beta)(\hat P_2-\beta)
{\rm sinh}^2 \delta_1\right ] - \cr
& & \left.\left [P_1P_2 (\hat Q_2+\beta)+ Q_1Q_2 (\hat P_1+
\beta)\right]^2 {\rm cosh}^2 \delta_1 {\rm sinh}^2 \delta_1 
\right |^{{1\over 2}},
\label{ent}
\end{eqnarray}
where again $\Delta \equiv {\rm sign}(Q_2) \sqrt{(Q_2)^2 
{\rm cosh}^2 \delta_1 - (P_1)^2 {\rm sinh}^2 \delta_1}$ and $\hat
P_1=+\sqrt{P_1^2+\beta^2}$, etc.  The entropy increases with 
$\delta_1$, approaching infinity [finite value] as $\delta_1 \to 
\infty$ [non-supersymmetric extreme limit is reached].  
For the supersymmetric extreme solutions, the entropy is 
\noindent{($a$) non-zero and finite, approaching infinity as 
$\delta_1 \to \infty$, when $P_{1,2}$ and $Q_{1,2}$ are non-zero, and}
\noindent{($b$) always zero when at least one of $P_{1,2}$,  
$Q_{1}$ (and $Q_2$) is zero with $\delta_1 \neq 0$ 
(with $\delta_1 =0$).}  

The Hawking temperature 
$T_H = |\partial_r \lambda (r=\beta)|/4\pi$ is given by
\begin{equation}
T_H={\textstyle {\beta\over \sqrt 2}} S^{-1}. 
\label{temp}
\end{equation}
As the boost parameter $\delta_1$ increases, the temperature $T_H$ 
decreases, approaching zero temperature.   In the supersymmetric 
extreme limit with at least three of $P_{1,2},Q_{1,2}$ parameters 
non-zero, the temperature is always zero.  With two of them 
non-zero, the temperature is non-zero and finite, approaching 
zero as $\delta_1 \to \infty$.  When only one of them is non-zero 
(only $Q_2$ is non-zero for the case $\delta_1 \neq 0$), the 
temperature becomes infinite.  In the non-supersymmetric extreme 
limit, the temperature is zero.

For the most general solution, it should be possible to write 
both the entropy (\ref{ent}) and the temperature (\ref{temp}) 
in the $T$- and $S$-duality invariant form, in terms of 
conserved (quantized)  28 electric charge  $\vec{\alpha}$ and 28 
magnetic charge $\vec{\beta}$ vectors, the asymptotic values of 
the moduli matrix $M$ (cf. (\ref{modulthree})), and the 
dilaton-axion fields $\varphi$ and $\Psi$.  In this case one 
would be able to address the dependence of  the above thermal
quantities on the asymptotic values of moduli and the couplings
\footnote{Note that in the supersymmetric limit, the 
entropy (\ref{ent}) has been cast \cite{CT} in such a form and 
is {\it independent} of the the moduli and couplings, as anticipated 
in Ref. \cite{LarsenWIl}.  We notice that for non-extreme solutions 
($\beta >0$) this is not the case anymore, and the entropy depends 
on the asymptotic values of the scalar fields.}.  

\section{Conclusions} 

In this paper, we obtained the most general static spherically 
symmetric solutions (compatible with no-hair theorem) of the 
effective four-dimensional heterotic string theory compactified 
on a six-torus, by performing a subset of $O(8,24)$ transformations
on the Schwarzschild solution.   Here $O(8,24)$ is the symmetry of 
the effective three-dimensional action of heterotic string, 
suitable for describing the stationary solutions of the 
four-dimensional heterotic string action.  
We provided an explicit sequence of symmetry transformations, 
which should be imposed on the Schwarzschild solution, in order 
to obtain the most general solution in this class. 

We gave the explicit form of the generating solution.  It was 
obtained by performing  a sequence of six $SO(1,1)\subset 
O(8,24)$ boosts on the Schwarzschild solution, with the zero 
Taub-NUT charge constraint imposing one constraint among two 
boost parameters.  The generating solution is therefore 
parameterized by six parameters:
the mass of the  Schwarzschild black hole and the six boost 
parameters,two of which subject to one constraint.   Equivalently, 
these six parameters can be traded for the ADM mass of the 
generating solution and three electric and three magnetic charges 
(subject to zero Taub-NUT charge constraint), which are
associated with the Kaluza-Klein and two-form gauge fields of  the
{\it two} compactified coordinates. The non-trivial scalar fields 
are the dilaton and the axion field, and the moduli of the two-torus. 

The general solution, which are parameterized by {\it unconstrained} 
28 magnetic  and 28 electric charges and the ADM mass compatible 
with the Bogomol'nyi bound, is obtained by imposing on the 
generating solution fifty $[SO(6)\times SO(22)]/[SO(4)\times 
SO(20)]\subset O(6,22)$ transformations and  one $SO(2)\subset 
SL(2,R)$ transformation. Here  $O(6,22)$ and $SL(2,R)$ are the 
$T$- and $S$-duality symmetries of the effective four-dimensional 
heterotic string action.  The above  subset  of $T$-duality 
transformations rotates the electric charges and like-wise
magnetic charges  as well as the  moduli fields, while  
the above $S$-duality transformation rotates each electro-magnetic 
charge combination and the dilaton-axion field.  Both subsets of 
transformations, however, do not affect the (Einstein frame) 
four-dimensional space-time, and thus the space-time properties 
of the whole class of solutions are determined by six parameters 
of the generating solution, which  in turn enabled one to analyze 
explicitly their global space-time structure and thermal properties.
  
The work also sets a stage for addressing the moduli and 
coupling dependence of the thermal quantities for the 
general class of static spherically symmetric configurations, 
which may in turn shed light on quantum aspects of black holes 
in string theory.

\acknowledgments
The work is supported by  the Institute for Advanced Study funds and
J. Seward
Johnson foundation, U.S. DOE Grant No. DOE-EY-76-02-3071, the
NATO collaborative research grant CGR No. 940870 and the National 
Science Foundation Career Advancement Award No. PHY95-12732.  
M.C. would like to thank F. Quevedo, A. Sen, A. Tseytlin and 
E. Witten for useful discussions.

\end{document}